\documentclass[sigconf]{acmart}

\usepackage{sidecap}
\usepackage{graphicx}
\usepackage[nodisplayskipstretch]{setspace}
\usepackage{booktabs} 

\usepackage[acronym,nomain]{glossaries}
\usepackage{enumitem}
\newlist{inparalist}{enumerate*}{1}
\setlist[inparalist]{label=(\roman*)}

\usepackage[compact]{titlesec}

\copyrightyear{2018} 
\acmYear{2018} 
\setcopyright{rightsretained} 
\acmConference[SIGSPATIAL '18]{26th ACM SIGSPATIAL International Conference on Advances in Geographic Information Systems}{November 6--9, 2018}{Seattle, WA, USA}
\acmBooktitle{26th ACM SIGSPATIAL International Conference on Advances in Geographic Information Systems (SIGSPATIAL '18), November 6--9, 2018, Seattle, WA, USA}
\acmDOI{10.1145/3274895.3274911}
\acmISBN{978-1-4503-5889-7/18/11}

\begin{document}

\title{Los Angeles Metro Bus Data Analysis Using GPS Trajectory and Schedule Data (Demo Paper)}

  \author{Kien Nguyen, Jingyun Yang, Yijun Lin, Jianfa Lin, Yao-Yi Chiang, Cyrus Shahabi}
\email{{kien.nguyen, jingyuny, yijunlin, jianfali, yaoyic, shahabi}@usc.edu}
\affiliation{%
  \institution{University of Southern California}
  \city{Los Angeles}
  \state{California}
  \postcode{90089}
}

%




\renewcommand{\shortauthors}{K. Nguyen et al.}

  \begin{abstract}
  With the widespread installation of location-enabled devices on public transportation, public vehicles are generating massive amounts of trajectory data in real time. However, using these trajectory data for meaningful analysis requires careful considerations in storing, managing, processing, and visualizing the data. Using the location data of the Los Angeles Metro bus system, along with publicly available bus schedule data, we conduct a data processing and analyses study to measure the performance of the public transportation system in Los Angeles utilizing a number of metrics including travel-time reliability, on-time performance, bus bunching, and travel-time estimation. We demonstrate the visualization of the data analysis results through an interactive web-based application. The developed algorithms and system provide powerful tools to detect issues and improve the efficiency of public transportation systems.
  \end{abstract}


\maketitle

    \section{Introduction}
    \newacronym{la}{LA}{Los Angeles}
    \newacronym{us}{US}{United States}
    \newacronym{gps}{GPS}{Global Positioning System}
  
  Inefficient traffic conditions have been a major issue for many big cities. The issue has led to significant cost of time and money for citizens and visitors. According to the Transportation Statistics Annual Report from the Bureau of Transportation Statistics in 2016~\cite{TSAR}, the average annual delay per commuter rose from 37 hours in 2000 to 42 hours in 2014, a 13.5\% increase. 
Traffic conditions in \gls{la} are even worse.  
	According to the TomTom Traffic Index, \gls{la} is ranked the most congested city in the \gls{us}, the 12th most congested city worldwide, with a typical half-hour commute taking 81\% longer during evening peak periods and 60\% longer during the morning peak.
    
    Fortunately, with the widespread establishment of public transportation in \gls{la}, traffic congestion can be alleviated. Also, in recent years, a vast number of location-enabled sensors (e.g., \gls{gps}) has been installed on buses running in \gls{la}, from which public transportation performance can be calculated and analyzed. The wealth of data collected from bus \gls{gps} trajectories offer an unprecedented opportunity for analysis of public transportation systems towards reducing operating costs and increasing ridership. Using bus sensor data from the past few years and transit schedule data publicly available from \gls{la} Metro, we are able to perform various kinds of analyses on the performance of \gls{la} metro buses.
    
    In this paper, we demonstrate algorithms and a system that process a massive amount of real-time and historical bus \gls{gps} trajectory datasets in \gls{la} County to analyze a variety of public transportation-system performance metrics. 
    In particular, the paper developed several novel components and algorithms to
	\begin{inparalist}
		\item clean and transform \gls{gps} datasets,
		\item map-match and integrate the cleaned datasets to the road network and timetables, and
		\item compute performance and reliability metrics such as travel-time reliability, bus bunching, and bus arrival-time estimation
	\end{inparalist}.
    The access and visualization of the performance metrics are enabled via a web-based application. 
    
    \subsection{Related Work}
    TRAVIC~\cite{bast2014travic} is a client for public transit movement visualization. However, it does not provide analytics from historical data.
    Lin et al.~\cite{lin1999experimental} presented an experimental system for bus arrival time prediction from \gls{gps} and scheduled data. However, their work focused on real-time prediction of arrival time of the next stop rather than analytics of historical estimated arrival time.
    TransDec~\cite{Demiryurek2010TransDecASQ} is a transportation decision-making system with wide range of data sources. However, it does also not provide analysis for buses. 

    \section{Definitions}
	\newacronym{id}{ID}{Identifier}
        In this section, we present definitions of variables used throughout the paper.
	\begin{definition}{\textit{Bus \gls{gps} records:}}
		A bus \gls{gps} record $p_i$ includes the route \gls{id} ($p_i.\mathit{route}$), an identifier with bus \gls{id} and run \gls{id} ($p_i.\mathit{identifier}$),  direction ($p_i.\mathit{dir}$), latitude and longitude ($p_i.\mathit{lat}$, $p_i.\mathit{lon}$), and time ($p_i.\mathit{time}$).
	\end{definition}
	\begin{definition}{\textit{\gls{gps} trip:}}
		A \gls{gps} trip is a sequence of consecutive \gls{gps} records $\tau = \{p_1, p_2, \dots, p_n\}$ that satisfy the property $\forall p_i. p_j \in P, p_i.\mathit{identifier} = p_j.\mathit{identifier}$
		and $\forall i \in [0, n-1], p_i.\mathit{time} < p_j.\mathit{time}$.
	\end{definition}
	\begin{definition}{\textit{Bus stop:}}
		A bus stop (or stop) $s_i$ includes stop \gls{id} ($s_i.\mathit{identifier}$), latitude and longitude ($s_i.\mathit{lat}$, $s_i.\mathit{lon}$).
	\end{definition}
    \begin{definition}{\textit{\gls{gps} run:}} 
    	a \gls{gps} trip without repeated stops.
    \end{definition}
	
	\begin{definition}{\textit{Scheduled Trip:}}
		A scheduled trip is a sequence of stop-times $\phi : \{st_1, st_2, \dots, st_n\}$ 
		indicating the progress of a \gls{gps} trip of a bus from one stop to the next stop. 
		Every stop-time $st_i$ has the route \gls{id} ($st_i.\mathit{route}$), expected arrival time ($st_i.\mathit{time}$) and a unique stop sequence ($p_i.\mathit{seq}$) 
		depicting the progress of the trip. 
		Each scheduled trip $\phi_i$ is also associated with a trip \gls{id} ($\phi_i.\mathit{trip}$) 
		and a service \gls{id} ($\phi_i.\mathit{service}$), which indicates days of the week that the trip would operate. 
	\end{definition}

    \section{Datasets}
	\newacronym{riits}{RIITS}{The Regional Integration of Intelligent Transportation Systems}
	\subsection{GPS Dataset}
	The \gls{gps} dataset is collected from \gls{riits}, a communication system that supports the real-time transformation of information sponsored by \gls{la} Metro. 
	Currently, 144 bus routes are operated and recorded by Metro.
	Approximately one million records are sent each day on average. 
	The time interval of each sensor data report is about 3 minutes.
	The main fields in the \gls{gps} dataset are route \gls{id}, bus \gls{id}, run \gls{id}, record time, latitude and longitude. 
	
	\subsection{Schedule Dataset}
	\newacronym{gtfs}{GTFS}{General Transit Feed Specification}
	The schedule dataset is provided by \gls{la} Metro in the \gls{gtfs} format with 38,719 trips of 146 bus routes in \gls{la}. 
	The \gls{gtfs} dataset used in this study is the December 2017 version, available on the LA Metro website.~\footnote{https://developer.metro.net/introduction/gtfs-data/download-metros-gtfs-data/}
	The original GTFS data is preprocessed into one integrated and dimensionally reduced dataset with only crucial information required for the performance metrics. 
	The processed dataset includes the following fields: corresponding route \gls{id}, trip \gls{id}, stop sequence, stop name, stop arrival time, latitude and longitude.

	\section{Data Preprocessing}
	\subsection{Overview}
	In this study, we have a set of individual \gls{gps} records that are not grouped into trips. In other words, there are only $p_{i}$ data but not $\tau_{i}$ data in the \gls{gps} dataset. Thus, to perform the mapping from bus \gls{gps} data to stops and schedules, we first have to group the \gls{gps} records into trips (i.e., $\tau_{i}$). Also, the bus direction data (i.e., $p_{i}.dir$) in the \gls{gps} data is not reliable. Thus, a bus direction recovery process is required.
	
\begin{figure}
		\centering
		\includegraphics[scale = 0.31]{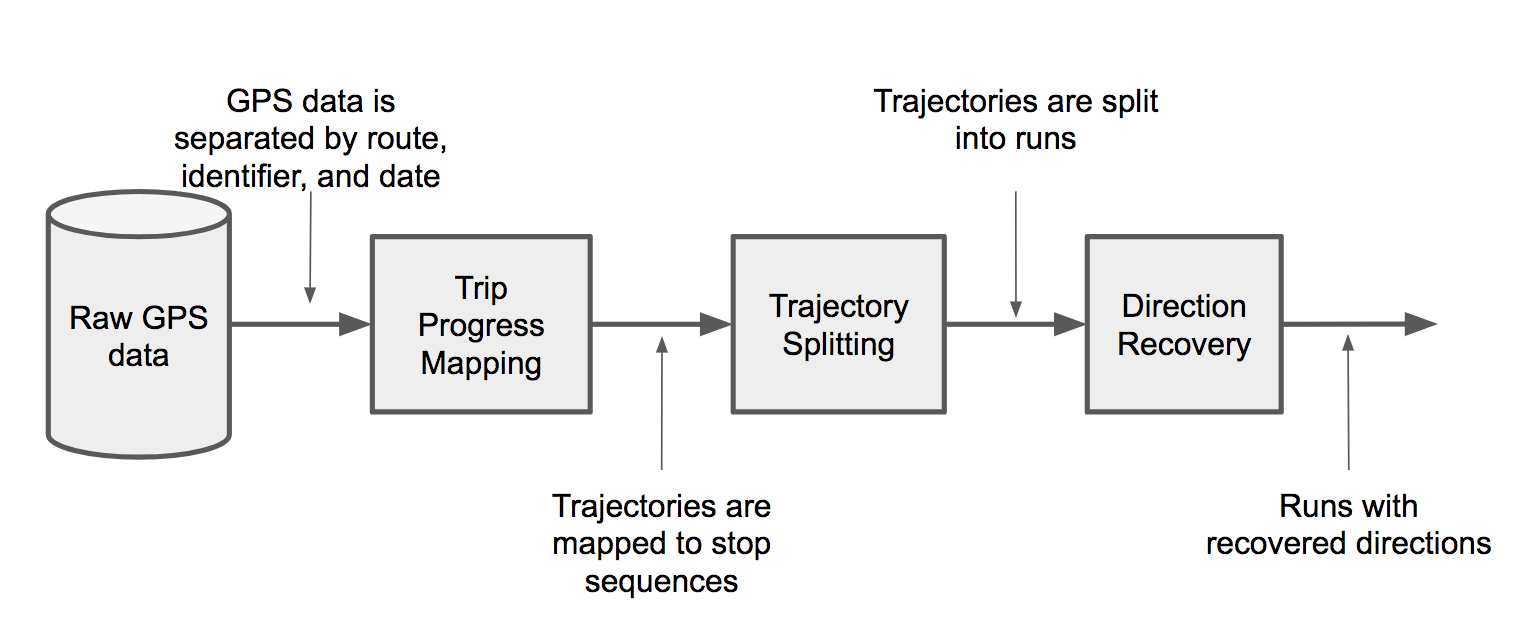}
		\caption{Pre-processing steps}
		\label{fig:preprocessing-steps}
	\end{figure}

	Figure~\ref{fig:preprocessing-steps} shows these data preprocessing steps. 
	First, raw \gls{gps} records, separated by route, identifier, and date, are mapped to stop sequences in the Trip Progress Mapping step. 
	Then, in the Trajectory Splitting step, the \gls{gps} records are split into individual trajectories where each trajectory represents a run of a bus from the start to end stations. The idle points are also removed from these trajectories by identifying the data points recorded when buses are not running or in intermission.
	Next, the direction of each record in each trajectory is recovered in the Direction Recovery step. The following sections describe these preprocessing steps in details.

	\subsection{Trip Progress Mapping}
	To split the records into different trips, we first need to know the bus positions in the trajectory. Thus, for the \gls{gps} records of the same route, identifier, and date, we have to extract the trip progress of buses from location data of the records. Our approach is based on the assumption that every valid trip in the dataset running the same route should travel a sub-trajectory of the route trajectory with a non-strictly increasing or decreasing stop sequence. 
	
	More formally, for every route $r'$, there exists a route trajectory $T'$ such that for every record $p_{i}$, $p_{i} \in P$, $r_{i} = r'$, there is a stop $t_{j} \in T'$ such that $D(p_i, t_j) < \epsilon$ where $\epsilon$ is a fixed value greater than 0. To perform the mapping, we select a parent trip of a route $r$ by selecting the trip with the longest cumulative distance for the route. Next, for every $p_{i}$ running route $r$, we define a mapping function $g: (lat,lon) \to seq$ that maps any location to the stop sequence of the record in the parent trajectory of the route that is closest to the location, and assign $g(p_i.lat,p_i.lon)$ to $p_{i}$. 
	Also, if the distance between the \gls{gps} point and the stop location is larger than a threshold (e.g., 400 meters), the \gls{gps} point is considered as an outlier and removed from the process.

	Therefore, after completing the trip progress mapping, every record $p_{i} \in P$ will have a stop sequence $p_i.seq$ associated with it depicting the buses' progress in the trip.

	\subsection{Trajectory Splitting}
	With bus trajectory splitting, we traverse through the set of \gls{gps} records with the same route \gls{id}, identifier, and running date. 
	For each record $p_i$ we associate a value $p_i.trend$ to the record such that:
	\begin{equation}
    \label{eq:trend}
	p_{i}.trend = sgn\left( g\left(p_{i}.lat,p_{i}.lon\right) - g\left(p_{i-1}.lat,p_{i-1}.lon\right) \right)
	\end{equation}
	After completing this step, the trend values of all the $p_i$'s are reviewed. If for a $p_i$, the largest integer $j$ s.t. $j<i; p_j.trend \neq 0$ and the smallest integer $k$ s.t. $k>i; p_k.trend \neq 0$ satisfies $ p_j.trend = p_k.trend $, then it means that $p_{j+1}, p_{j+2}, \cdots, p_{k-1}$ belong to the same trip as $p_{j}$ and $p_{k}$. Thus, $p_{m}$ is set to $p_j$ ($p_k$) for all $j \leq m \leq k$. After this step, every group of consecutive records with the same trend value that is not zero should belong to a trip. Thus, at this point, trips can be split into groups with consecutive records with the same trend value, which is called a ``run". 
    Groups with a zero value at the beginning or end of each run are deleted as they are identified as idle points.
	Then, all runs having too few \gls{gps} records (e.g., fewer than five records) are removed. 

	\subsection{Direction Recovery}
	For every run, there is a unique trend value associated with consecutive \gls{gps} points (Equation~\ref{eq:trend}). We then map the trend value to the representation of direction value in $p_{i}$. 
    
    \section{Arrival Time Estimation}
	In this section, our goal is to estimate the arrival time at each stop by combining the \gls{gps} data and schedule data.
	The reason is that most of the \gls{gps} points are not recorded when buses arrive at the stops.
	This is an essential step because arrival time of a bus to a stop can provide information to compare the estimated arrival time and the schedule, which in turn can help us calculate many different metrics of the data analytics process.
	
\begin{figure}
		\centering
		\includegraphics[scale = 0.22]{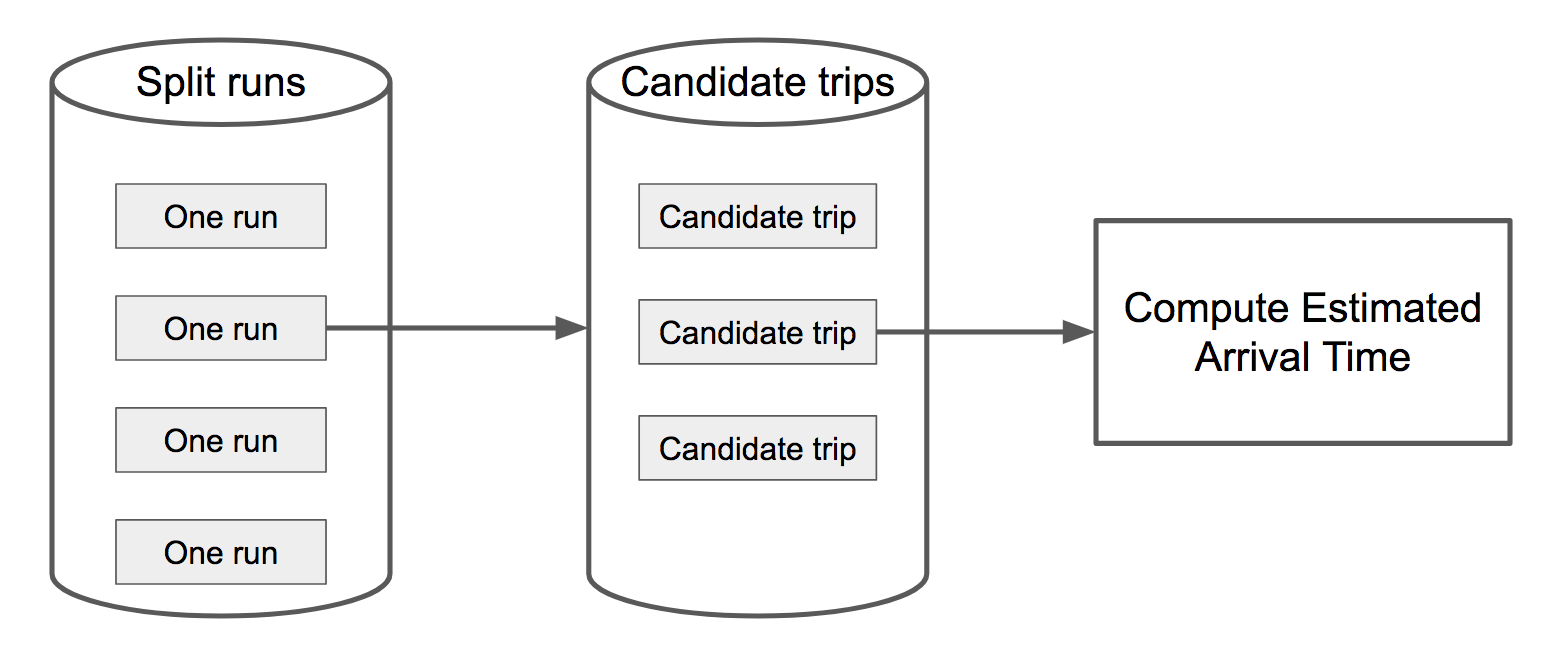}
		\caption{The steps to estimate arrival time}
		\label{fig:estimateArrivalTimeSteps}
	\end{figure}

	The main steps to estimate arrival time are shown in Figure~\ref{fig:estimateArrivalTimeSteps}.
	The general idea is that for each run, we find its candidate trips (schedules), and compute the estimated arrival time for the stops in each candidate of that run.
	
	First, for a run of a bus of a specific route, to reduce the number of candidate trips to which the run may belong, all the trips of that route with the start and end time covering the start and end time of the run are considered as candidate trips. After that, for each candidate trip, we calculate the distance from the \gls{gps} run to the trip. The distance is calculated based on the sum of the distances from each \gls{gps} record of the run to its closest stop of the trip.

	Next, the candidate trips are sorted by their distance to the \gls{gps} run, and the candidate trips with the distance smaller than a threshold compared to the closest candidate are selected. 
	Thresholding by distance is helpful because there can be several trips that share the same stops with different schedule times. 
	
	Subsequently, for each candidate trip of a \gls{gps} run, the arrival time of the bus for each scheduled time at a stop is estimated. The process includes finding the \gls{gps} records of the run that are the first records coming before and after the location of the stop, then calculating distance from the stop to those two \gls{gps} records, assuming that the bus ran at a constant speed between those two \gls{gps} records, and finally estimating arrival time based on distances and time at the \gls{gps} records.   
    
	\section{Data Analytics}
	\subsection{Travel-Time Reliability}
	Travel-time reliability (or on-time performance) indicates the percentage of seeing a bus at a scheduled time of a trip at a stop. Based on the definition from \gls{la} Metro,~\footnote{\url{https://www.metro.net/about/metro-service-changes/service-performance-indicators}} the range of one minute early and five-minutes late is used as the range for on-time achievement. 
	Thus, from estimated arrival times of buses to a specific schedule, a scheduled time at a stop is marked as on-time if there is an estimated arrival time of a bus lying within the range of on-time achievement. Finally, the percentage of on-time scheduled times are calculated for each route, for each stop on a route, and for each scheduled time (trip) of a stop of a route. 
	
	\subsection{Travel-Time Deviation}
	The travel-time deviation is the difference between a scheduled time of a trip at a stop and the estimated arrival time of a bus based on \gls{gps} data. 
	All the values of the travel-time deviation are then averaged to show the average travel-time deviation.

	\subsection{Bus Bunching}
	Due to variety of reasons such as traffic jams or large crowd of passengers, some buses may run slower than other buses, thus creating the situation where there are several buses of the same route running too close to each other. This situation is called \textit{bus bunching}. 
	
	Therefore, bus bunching can be calculated as the number of buses arriving within the one minute early and five-minute late range at a scheduled time of a trip at a stop. 
	If there is more than one arrived bus, a bus bunching event is marked. Consequently, the percentage of bus bunching can be calculated.

	\subsection{Travel-Time Estimation (Waiting Time Estimation)}
	Travel-time (or waiting time) estimation shows the average time a passenger has to wait for the bus to arrive at the scheduled time of a trip at a stop. This estimation can also be calculated from the estimated arrival times by choosing the minimum positive value of delay time as the time a passenger has to wait. Consequently, the average travel-time estimation can be computed.

	\section{Demonstration}
	\subsection{System Architecture}
\begin{figure}
		\centering
		\includegraphics[scale = 0.1]{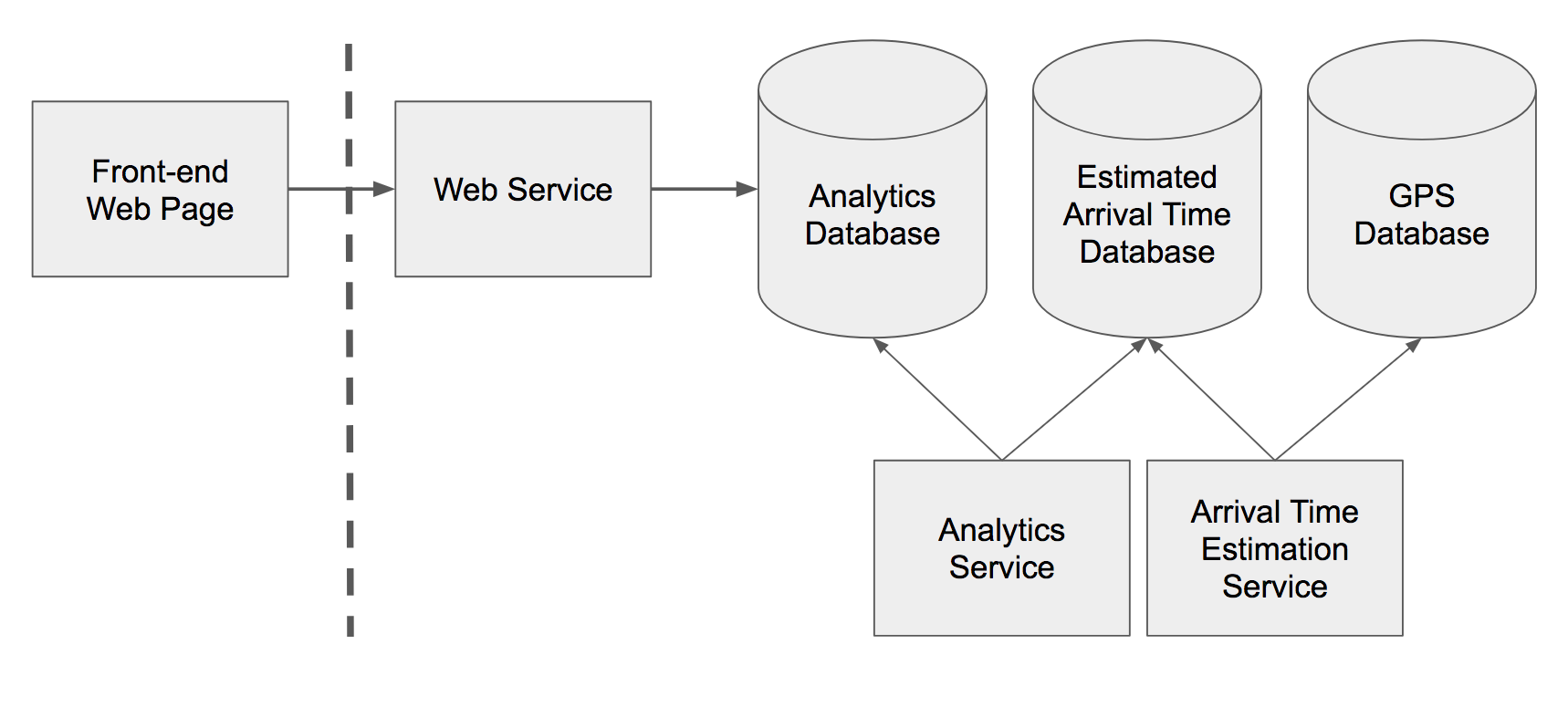}
		\caption{System Architecture}
		\label{fig:systemoverview}
		\vspace*{-5pt}
	\end{figure}

	The system architecture for this demonstration is shown in Figure~\ref{fig:systemoverview}. 
    The \gls{gps} Database stores all \gls{gps} data. 
    The Arrival Time Estimation Service fetches \gls{gps} data from the \gls{gps} database, estimates arrival times, and stores them into the Estimated Arrival Time Database.
    The Analytics Service uses the estimated arrival times to calculate different analytics such as travel-time reliability, travel-time deviation, bus bunching, and travel-time estimation. 
    The analytics are stored in the Analytics Database and can be fetched by the Web Service to serve data to other agencies. In the next section, different parts of the system are explained. 
    A combination of Oracle and PostgreSQL databases are used for data storage; the services are written in Java; and the web application is written using React framework.
    The web application can be accessed at \url{http://gdt.usc.edu/METRANS/dashboard/}.
    
    \begin{figure*}
		\centering
		\includegraphics[width=0.7\textwidth]{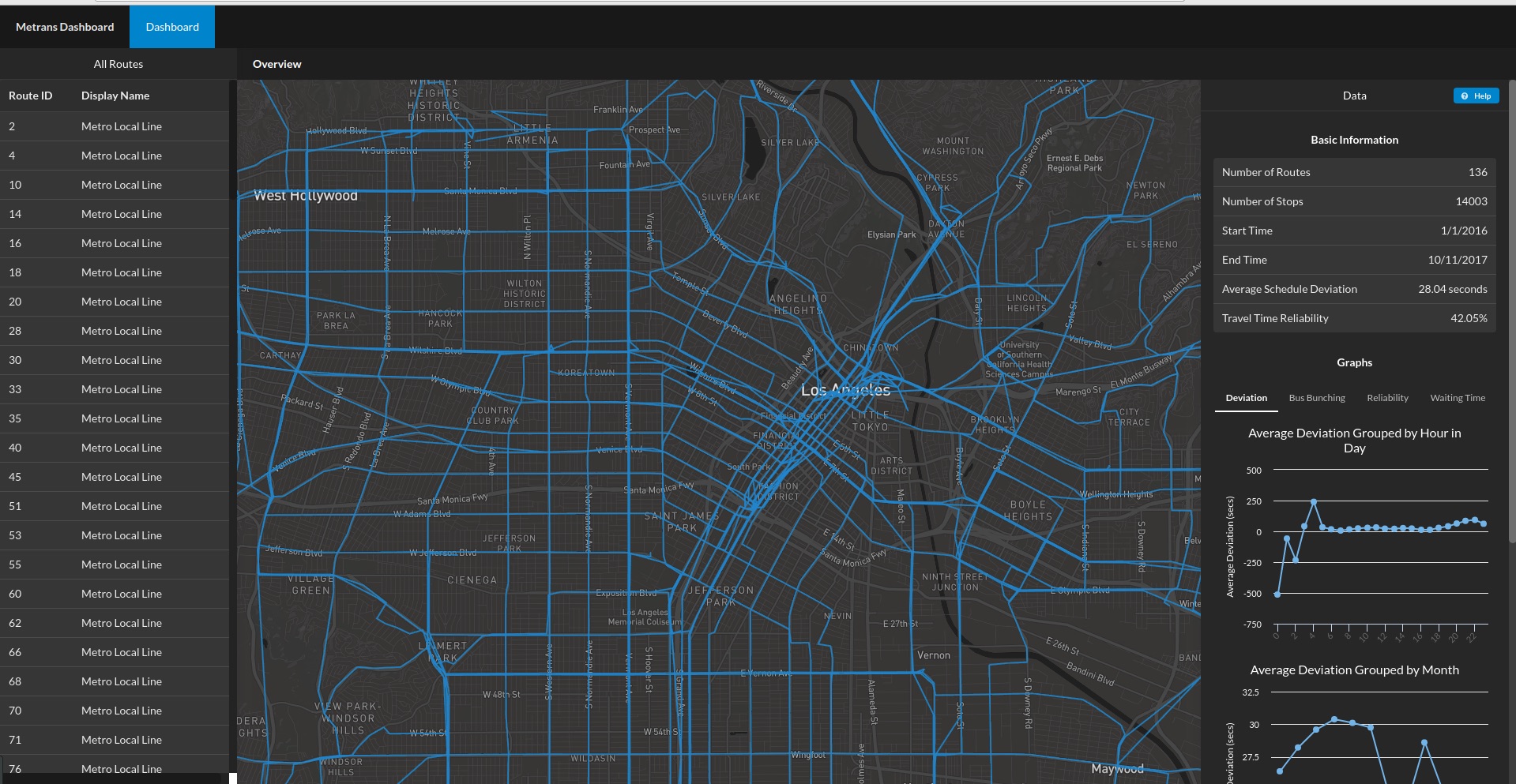}
		\caption{An example page of the web application. On the left is the list of all bus routes; in the center is the map with the shape of routes displayed; and on the right are the analytics}
		\label{fig:dashboard-overview}
	\end{figure*}
    
	\subsection{Overview Page}
	The Metrans Dashboard is the web front-end for users to interact with analyses.
	When a user clicks on the ``Dashboard" button on the top-left corner, the user is navigated to the Overview page (Figure~\ref{fig:dashboard-overview}). 
	In the Overview page, the list of all bus routes is shown on the left to provide more information about a specific route. In the center is the map with the shape of routes, and on the right is the overall information of all routes. 
	
	The ``Basic Information" box shows basic information of our dataset, such as the number of bus routes or the number of stops. 
	The box also shows overall information such as the average travel-time reliability over all routes is about 42\%. 
	Below the box are various graphs showing our measurements, which are Deviation, Bus Bunching, Reliability, and Waiting Time, grouped by the hour of the day, the day of the week, and the month.

	\subsection{ Route-Specific Page}
	When the user clicks on a specific route, the user is navigated to a route-specific page which shows more information of the route.  
    The left side shows the list of all stops of the route, the shape of the route is displayed on the map in the center, and more analyses are shown on the right. 
	

	\subsection{Stop-Specific Page}
\begin{figure}
		\centering
		\includegraphics[scale = 0.11]{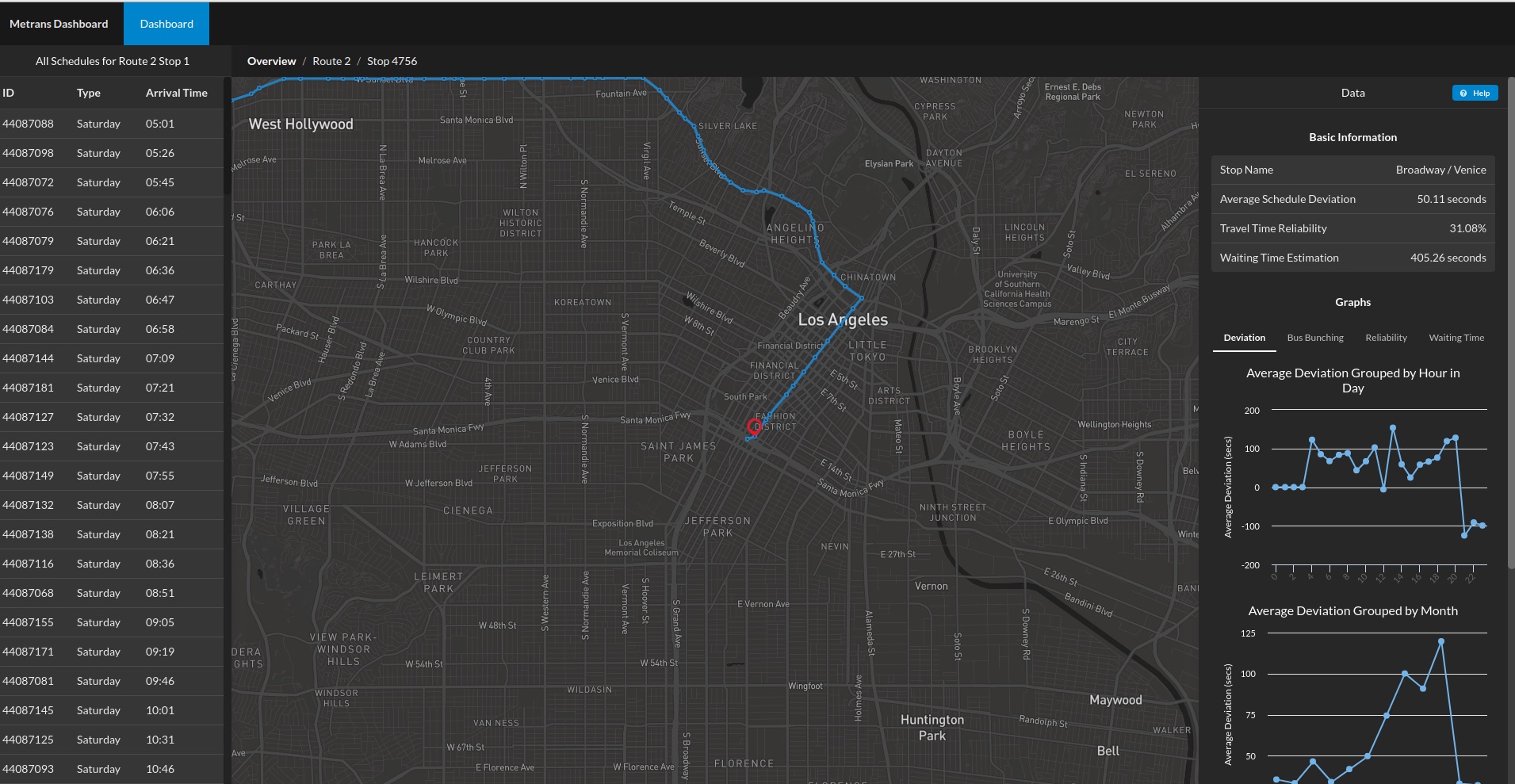}
		\caption{Stop-Specific Page}
		\label{fig:dashboard-stop}
	\end{figure}

	On a route-specific page, the user can click on a specific stop, and will be directed to a stop-specific page for that stop. Figure~\ref{fig:dashboard-stop} shows the stop-specific page when the user clicks on Stop 4756 of Route 2. 
The left side shows the list of all scheduled trips for Stop 4756 of Route 2 with the Service days (e.g., Saturday, Weekday) and the scheduled arrival of the trip. In the center is the location of the stop marked on the map, and the right side is more  statistics for the stop. 
	
	\subsection{Trip-Specific Page}
	When the user clicks on a specific trip of a stop, the stop-specific page changes the information of the stop on the right side to the information of the selected-trip, which is highlighted on the list. 
	
	
    \section{Discussion and Future Work}
    This paper presented the algorithms and a system that support the analysis of various public transportation performance metrics from real-time GPS and scheduled data in \gls{la} County. We plan to develop machine learning algorithms for forecasting the presented performance metrics and incorporate the algorithms into the current system. 

    \vspace*{-5pt}
    \section*{Acknowledgement}
    This research has been funded in part by NSF grant CNS-1461963, Caltrans-65A0533, LA Metro contract LA-Safe-PS36665000, the USC Integrated Media Systems Center, the USC METRANS Transportation Center. Any opinions, findings, and conclusions or recommendations expressed in this material are those of the authors and do not necessarily reflect the views of any of the sponsors such as NSF.

    \vspace*{-5pt}
    \bibliographystyle{ACM-Reference-Format}
    \bibliography{sample-bibliography}

\end{document}